# Intrinsic structural features in random packings of monodisperse spheres

S. Q. Jiang and M. Z. Li*

*Department of Physics, Beijing Key Laboratory of Opto-electronic Functional Materials & Micro-nano Devices, Renmin University of China, Beijing 100872, China*

**Abstract**

Unveiling the characteristic structural features of random packings is a long-standing challenge. In this work, we developed a new tetrahedral structure model including not only regular tetrahedron (T), but also quartoctahedron (Q) and BCC simplex (B). The geometrical structure of jammed random packings of monodisperse spheres with different friction coefficients was systematically characterized. An intrinsic structure feature is revealed for all jammed random packings, which is determined only by the packing fraction, independent of interparticle friction. With increasing packing fraction, the population of T, Q, and B increases. Moreover, it is found that the structural configurations associated with crystal phases formed by face-adjacent T, Q and B increases as packing fraction increases, which is more prevalent than the geometric frustrated ones formed by face-adjacent T. In addition, it is revealed that instead of the regular simplexes of T, Q, and B, the irregular simplexes play more important roles in the density increase in disordered packings. Our findings provide new understanding for the structural nature of disordered packings and the underlying structural basis of the random close packing.

* maozhili@ruc.edu.cn

## I. Introduction

Finding the most efficient way to pack spheres is among the oldest puzzles known to scientists [1,2]. Apart from the mathematical significance [3-5], packings of spheres have been widely applied to the structural studies of living cells [6] and liquids [7-12], colloids [13,14], glass transition [15-19], phase transition between disordered and ordered states [20-29], and jamming system [30-36]. For the simplest case, the most efficient way to pack identical spheres has been studied extensively. In 1611, Kepler conjectured that the densest packing of spheres could be achieved by stacking close-packed planes with packing fraction of $\pi/\sqrt{18} \approx 0.7405$, which was proved only recently [4]. For the more complex case, the random sphere packings were first explored by Bernal in 1960's. He found that the most compact way to randomly pack spheres, that is, random close packing (RCP), results in a maximum packing fraction of $\phi_{RCP} \sim 0.64$, beyond which packings are found to contain crystalline regions inevitably [8-10,21,25,29,35]. Moreover, recent theoretical studies on the jammed random packings suggest that RCP state with the value of $\phi_{RCP}$ of the frictional hard spheres changes by varying interparticle friction coefficient from zero to infinity [33]. Thus, this indicates that RCP is not a unique point, but an RCP line from the frictionless point (RCP point) to the point with infinite friction coefficient. Although $\phi_{RCP}$ seems to be robust and of great physical significance, its nature remains highly controversial [5,8,21,22,24,27,31,32,35,37-42].

From the thermodynamic perspective for the repulsive potential jam, it is found that the jamming threshold is very close to 0.64 [32]. By developing a mean-field approach of jammed random packings, the states lying in the RCP line can be interpreted as the ground states of hard spheres characterized by different interparticle friction coefficients [33]. Moreover, the entropy of the jammed random packings was found to decrease with increasing packing fraction, indicating that the states in the RCP line is more ordered than other random packings [34,43]. Although these studies enriched the understanding of RCP, no detailed structure feature in disordered sphere packings was characterized, which is crucial for understanding the nature of RCP and structural characteristics of disordered packings.

Recently, by investigating tetrahedral configurations of frictionless hard spheres, polytetrahedral aggregates formed by face-adjacent tetrahedra are revealed to increase with packing fraction and become the most prevalent at RCP point [21,24]. This indicates that the geometric frustrated polytetrahedral configurations are the essential feature of random sphere packings and responsible for the densification as packing fraction increases [21,24]. Polytetrahedral model provides a plausible structural perspective for the nature of RCP and densification mechanism in random packings. However, it cannot explain how the crystal nuclei formation starts in the disordered packings with such a geometric frustrated polytetrahedral configuration filled in everywhere, as the density passes through the critical density $\phi_{\text{RCP}}$. Therefore, the polytetrahedral configurations could be only one aspect of the structural characteristics of random packings. To obtain more comprehensive structural features, careful characterization and classification of tetrahedral configurations in random packings are more important.

Considering the tetrahedral features of face-centered cubic (FCC), hexagonal close-packed (HCP) and body-centered cubic (BCC) crystals [44], in this work, we developed a new tetrahedral structure model for characterizing the geometric structural features of jammed random packings of monodisperse spheres with different friction coefficients. In this model, not only tetrahedron (T), but also quartoctahedron (Q) and BCC Simplex (B) were taken into consideration. It is revealed that the fraction of T, Q, and B increases with increasing packing fraction, and depends only on packing fraction, indicating that there is intrinsic structure feature in disordered sphere packings. It is also found that instead of geometric frustrated configurations, the structural configurations associated with crystal phases are much more prevalent and increase as packing fraction increases, which provides natural structural basis for understanding the mechanism of the disorder-to-order transition at $\phi_{\text{RCP}}$.

## II. Structure Model

In this work, the random packings of monodisperse spheres with different friction coefficients were investigated, and the simulation details can be found in Appendix A. In order to thoroughly characterize the structural features in random packings of monodisperse spheres, we developed an extended tetrahedral structural model, in which tetrahedron (T), quartoctahedron (Q) and BCC Simplex (B) were taken into consideration, as shown in FIG. 1(a-c), because both FCC and HCP crystal structures consist of T and Q, and BCC structure consists of BCC simplexes, as shown in FIG. 1(d-f). In FCC structure, every T is adjacent by faces only to four Q, and every Q is face-adjacent to two T and two Q, so that there are two types of clusters in FCC structure, as shown in FIG. 1(g). Similarly, there are three types of clusters in HCP structure, as shown in FIG. 1(h). In BCC structure, however, there is only one type of cluster formed by five B with face-adjacent to the central one shown in FIG. 1(i). Thus, one can characterize the local configurations associated with crystalline order by analyzing the above crystal-type clusters which contains only 8 spheres, as shown in FIG. 1(g-i) [8,21,24,44].

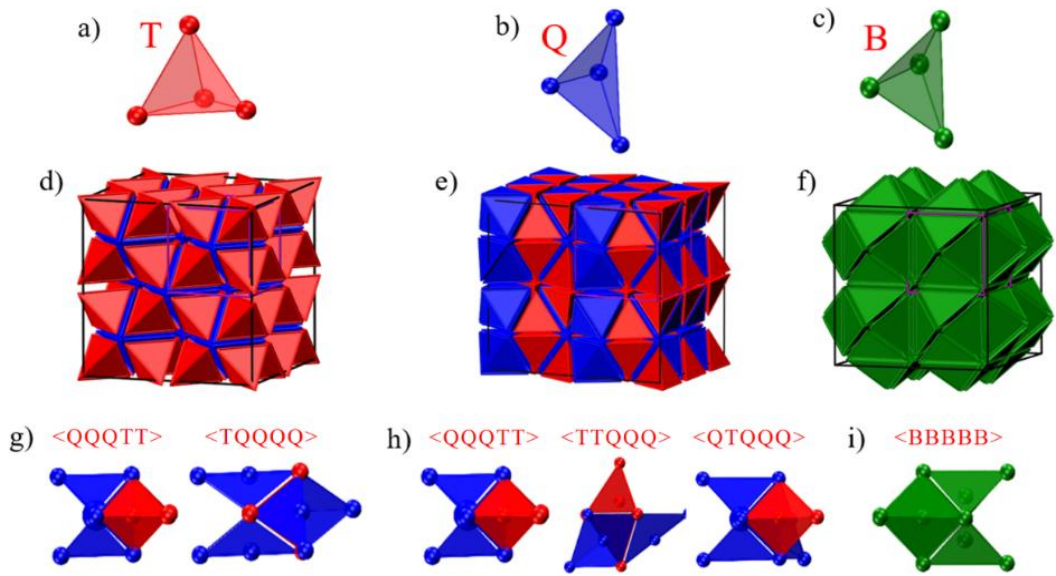

**FIG. 1. The structure model. (a-c)** The three basic building blocks of tetrahedron (T), quartoctahedron (Q), and BCC Simplex (B) in FCC, HCP, and BCC crystal structures, respectively. **(d,e)** Both FCC and HCP structures consist of tetrahedron and quartoctahedron. **(f)** BCC structure consists of only BCC Simplex. **(g-i)** Local clusters formed by five building blocks with four neighbors adjacent to a central one by faces in FCC, HCP, and BCC structures, respectively. **(g)** In FCC crystal, there are two types of clusters denoted by <QQQTT> and <TQQQQ>, respectively. Here the first letter represents the central building block, and the others represent 4 neighbors adjacent to the central one by faces. **(h)** In HCP crystal, there are three types of clusters: <QQQTT>, <TTQQQ> and <QTQQQ>. **(i)** In BCC crystal, there is only one type of cluster: <BBBBB>.

In addition, in these crystal-type clusters, if one or two neighboring simplexes do not satisfy the crystal-type cluster configurations, such clusters are classified as defective clusters and denoted as defective I and defective II, respectively. Moreover, this model can also extract the non-crystal local structures with T having at least two face-adjacent T in its neighborhood, and the other neighboring simplexes can have an arbitrary shape in general [8,21,24,44]. Since the arrangements of more than two T are incompatible with translational symmetry, they are geometric frustrated local structures. Polytetrahedral clusters formed by face-adjacent T and five-membered rings of tetrahedra (pentagonal bipyramids) can be identified by this structural type. Here Procrustean distance approach was employed to evaluate the shape of an arbitrary Delaunay simplex in these disordered packings and identify T, Q, and B (see Appendixes B&C for details).

## III. Results

To reveal the general structural features dependence on the density, FIG. 2(a-c) show the volume fraction of T, Q, and B as a function of packing fraction $\phi$ and mechanical coordination number Z for the jammed random packings, respectively. It can be seen that the volume fraction of T, Q and B increases continuously with increasing packing fraction, from about 3%, 7% and 4% at $\phi \approx 0.54$ to about 14%, 20% and 11% at $\phi \approx 0.64$, respectively. This indicates that the structures of random packings become more and more regular with increasing packing fraction. In addition, as packing fraction is close to $\phi_{\text{RCP}}$, the total volume fraction of T, Q, and B reaches more than 40% (see Supplementary Figure S1), which implies that these regular simplexes may play important role in the disorder-to-order transition as the packing fraction passes through $\phi_{\text{RCP}}$.

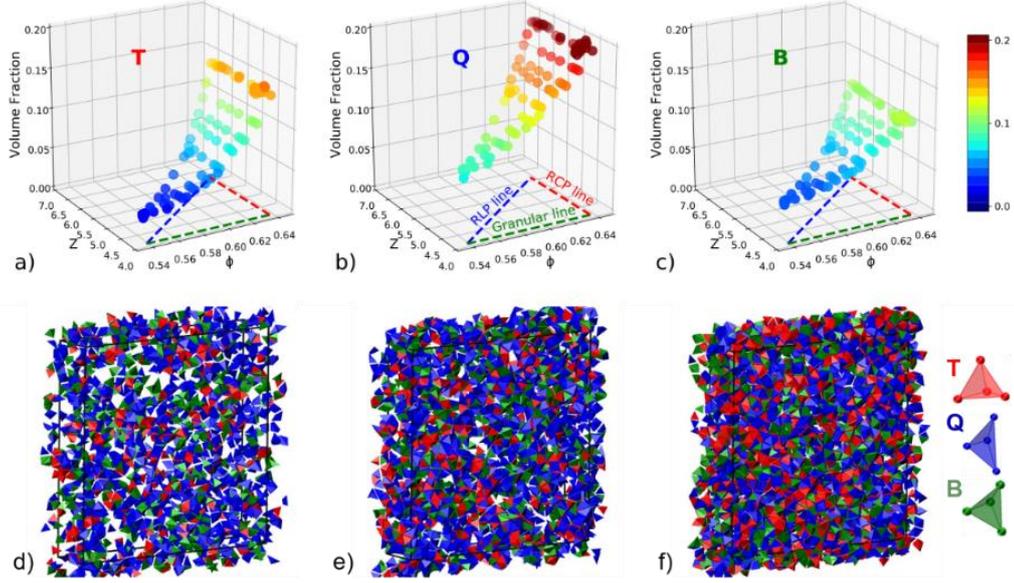

**FIG. 2. Population of the regular simplexes in disordered packings.** The volume fraction of T (a), Q (b) and B (c) at different packing fractions $\phi$ and mechanical coordination number Z. The color of spheres indicates the volume fraction. All jammed disordered packings lie within the triangle demarcated by the dashed lines in $Z - \phi$ plane which represent the random loose packing (RLP) line, random close packing (RCP) line, and granular line, respectively [33]. The space distribution of regular simplex (T, Q and B) at $\phi = 0.539, 0.603, 0.636$ show in (d), (e) and (f) respectively. The size of box slice is 0.2x * y * z.

Interestingly, it is observed in FIG. 2(a-c) that the population of the regular simplexes of T, Q, and B is almost the same for the jammed random packings with the same packing fraction, independent of mechanical coordination number Z. This reveals that the random sphere packings possess intrinsic geometric structural features, which is independent of interparticle friction or preparation protocol. Previous studies show that the volume fluctuation characterized by Voronoi volume is the same for random packings at given packing fractions, independent of friction coefficients [34,43]. Our results reveal the underlying intrinsic structural basis for the disordered sphere packings. This finding also provides a new understanding for the nature of RCP. Although the random sphere packings at RCP line keep the disordered feature without crystal phases, they may contain the regular simplexes as many as possible compared to those below $\phi_{\text{RCP}}$. In this sense, the randomness of RCP may be well-defined based on the above structure analysis.

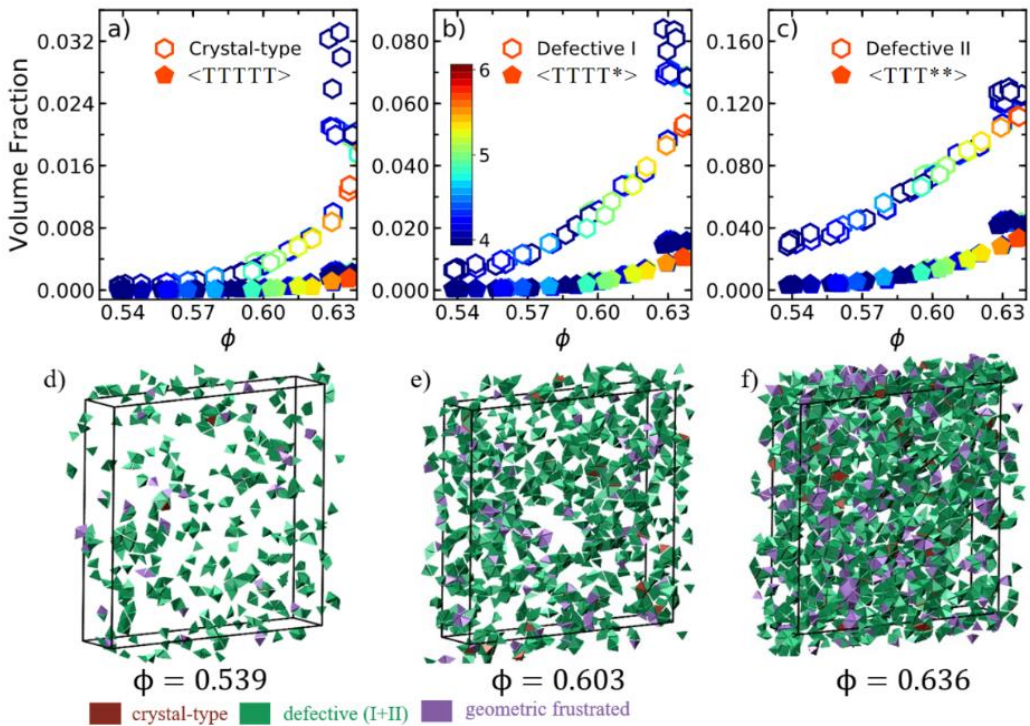

**FIG. 3. Population of crystal-related and geometric frustrated clusters in disordered packings. (a-c)** Comparison of the volume fraction of the crystal-type, defective (I and II), and geometric frustrated clusters in the jammed disordered packings with different packing fraction $\phi$ and mechanical coordination number Z. The color of each symbol represents Z value indicated by the color bar in (b). Here <TTTTT>, <TTTT*> and <TTT**> denote the geometric frustrated clusters, where * represents a simplex with arbitrary shape except T. **(d-f)** Spatial distribution of the central simplexes in the crystal-type, defective (I and II), and geometric frustrated clusters at $\phi = 0.539$, $\phi = 0.603$, and $\phi = 0.636$, respectively.

To reveal the local structural features, FIG. 3(a-c) shows the volume fraction of the central simplexes in crystal-type, defective I, defective II and geometric frustrated clusters as a function of packing fraction for the jammed random packings with different Z, respectively. This indicates that both crystal-related clusters (including crystal-type, defective I and defective II clusters) and geometric frustrated clusters increase with increasing packing fraction. However, the population of the crystal-related clusters is much higher than the geometric frustrated ones, and increases more quickly. Therefore, the crystal-related clusters dominate the local ordering in all random sphere packings. The population of each type cluster in the packings lying in the RLP line can be found in Supplementary FIG. S2. In specific, FCC- and HCP-type clusters increase more significantly. This indicates that the local configurations associated with FCC and HCP crystal phases may play much more important roles than the geometric frustrated structures in the disorder-order transition as packing fraction goes beyond $\phi_{RCP}$ [26,39]. The spatial distribution of the central simplexes of the crystal-type, defective, and geometric frustrated clusters in the packings at RLP line is also illustrated in FIG. 3(d-f). It is clearly seen that the clusters associated with crystal phases are more prevalent in random packings, with the geometric frustrated clusters embedded in the crystal-related clusters. This is more significant at RCP point shown in FIG. 3(f).

We also characterized the polytetrahedral configurations formed by three or more face-adjacent tetrahedra in random sphere packings according to our structural model, as shown in FIG. 4. The volume fraction of polytetrahedral aggregates characterized in our model is only about 7.5% at $\phi_{RCP}$, while the volume fraction of the crystal-related clusters is more than 20%. The above results demonstrate that the configurations associated with crystal phases are more important than the geometrical frustrated ones in disordered packings as packing fraction passes through the RCP density. Such structure feature in the disordered packings at RCP line can naturally explain the crystal nuclei formation and disorder-to-order transition at $\phi_{RCP}$. This can be further understood by exploring the densification mechanism in disordered packings within our structure model.

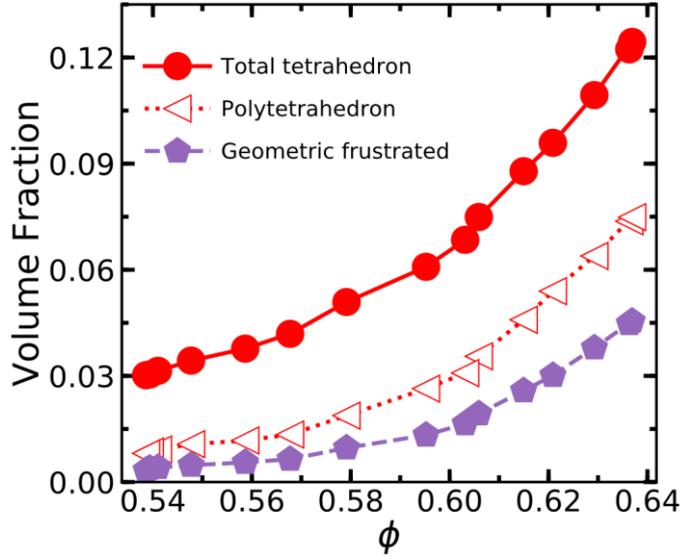

**FIG. 4. The volume fraction of the Tetrahedra, polytetrahedra and the geometric frustrated tetrahedra as a function of packing fraction in disordered packings lying in the random loose packing (RLP) line.** Here polytetrahedra is tetrahedron cluster formed by three or more face-adjacent tetrahedra. For each geometric frustrated tetrahedron, there are at least two tetrahedrons adjacent to it by face, which exclude the tail tetrahedral. Therefore, not all tetrahedra in polytetrahedral are geometric frustrated.

Previous studies assume that the increase of T is responsible for the increase in the density of random packings below $\phi_{RCP}$, because of its rather high local density [21]. However, no quantitative analysis was derived to confirm it. On the other hand, the population of T shown in FIG. 2(a) is not high, only about 14% even at $\phi \approx 0.64$. Thus, the increase of the number of T might not be able to cover the density increase of random packings. To understand the above contradiction and get more insights into the mechanism of density increase in random packings with increasing packing fraction, it is quite useful to directly analyze the mean local packing fraction of different types of simplexes in random packings, so that the density increase of each type of simplexes can be calculated. The packing fraction of a simplex can be defined by the ratio of the occupied volume by four spheres in the simplex to its total volume [45]. The mean packing fraction of one type of simplexes is given by the harmonic mean method [28,29]. For example, the mean packing fraction of T can be expressed as $\phi_T = \frac{N_T}{\sum_{i=1}^{N_T} 1/\phi_T^i}$, with $\phi_T^i$ and $N_T$ being the packing fraction of $i$th T and the total number of T, respectively. As shown in FIG. 5(a), the mean packing fraction of all types of simplexes increases as packing fraction increases. Thus, all types of simplexes become denser and denser, making contributions to the density increase. It can be also seen that the mean packing fraction of T, Q, and B is more than 0.6 even in the packing at RLP point, much higher than irregular simplexes, showing the nature of dense packing in T, Q, and B. Although the mean packing fraction of T, Q, and B continuously increases, the increase rate is getting slower, and the mean packing fraction becomes almost saturated far below $\phi_{\text{RCP}}$, especially for T and Q. As shown in FIG. 5(a), the mean packing fraction of T, Q, and B changes very little as $\phi > 0.6$. In contrast, however, the mean packing fraction of irregular simplexes increases linearly and exceeds 0.6, as packing fraction $\phi$ approaches $\phi_{\text{RCP}}$. Therefore, the density increase in irregular simplexes is more significant than that in T, Q, and B. Since the volume fraction of the irregular simplexes is much larger, the irregular simplexes may play more important role in the density increase in random packings.

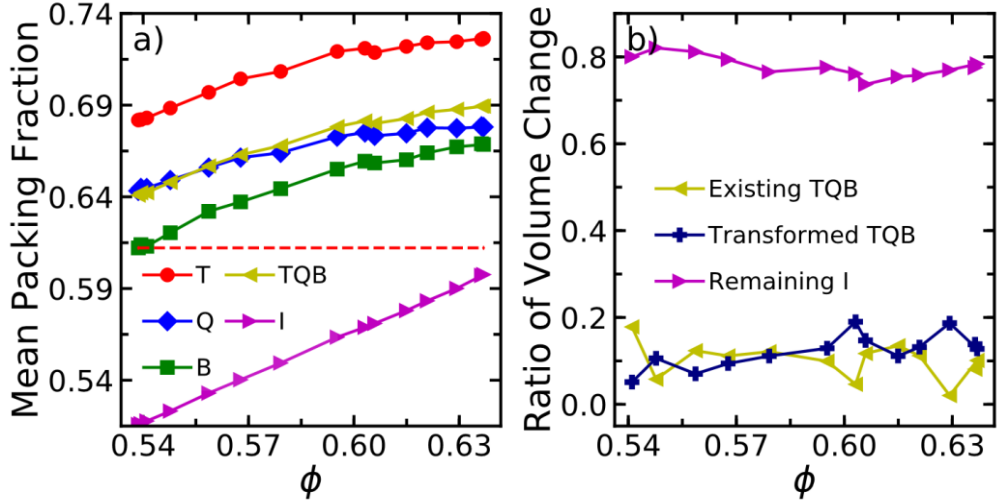

**FIG. 5. Mechanism of increase in density of disordered packings.** (a) Mean packing fraction of T, Q, B, and irregular simplexes (I) as a function of packing fraction. The mean packing fraction of the regular simplexes (T+Q+B) was also presented. The horizontal dashed line marks the mean packing fraction of BCC Simplex at random loose packing (RLP) point ($\phi = 0.539$). (b) The ratio of volume change to the total volume change between two adjacent disordered packings for the existing regular simplexes, newly formed T, Q, or B, and the remaining irregular simplexes, respectively.

To more precisely evaluate the contribution of regular and irregular simplexes to the density increase of random packings, we analyzed the volume change of each simplex between two adjacent packings of $\phi_1$ and $\phi_2$ ($\phi_1 < \phi_2$), and classified them into three groups: (1) the volume change of the existing regular simplexes as packing fraction increases from $\phi_1$ to $\phi_2$, (2) the volume change due to the change of the irregular simplexes in $\phi_1$ to regular ones in $\phi_2$, (3) the volume change of the remaining irregular simplexes in $\phi_1$ which are still irregular ones in $\phi_2$. Since the disordered packings used in this work were generated independently, only the number of simplexes in each group was considered. FIG. 5(b) shows the ratio of volume change in three groups to the total volume change between two adjacent packings. Surprisingly, the existing and newly formed regular simplexes contribute only about 20% to the density increase. In contrast, however, the remaining irregular simplexes make contributions of about 80%. This is in contrast to the previous results that an increase in the number of tetrahedral configurations is responsible for the increase in the density of a random packing before $\phi_{RCP}$ [21,24,44]. Our results elucidate that instead of regular simplexes (T, Q, and B), the irregular simplexes dominate the increase in the density, suggesting a new mechanism in density increase in random packings. In general, all simplexes in random packings become denser as packing fraction increases. However, the density increase in the regular simplexes of T, Q, and B is very limited, especially in the disordered packings of $\phi > 0.6$, because they are already very dense. Although more T, Q, and B are formed with increasing packing fraction, their increase between two adjacent packings is also very limited, which cannot account for the density increase, either. In contrast, irregular simplexes are more loosely packed, and their population is much higher, compared to the regular ones. Therefore, they can easily change shapes and become more regular and denser with increasing packing fraction, making greater impact on the density increase. This can be further elucidated in terms of the local packing fraction distribution of different types of simplexes, as shown in FIG. 6. The distribution of irregular simplexes as well as all simplexes shifts to higher density and becomes much narrower with increasing packing fraction. Our results suggest that the density increase in random packings relates to a continuous change in structures.

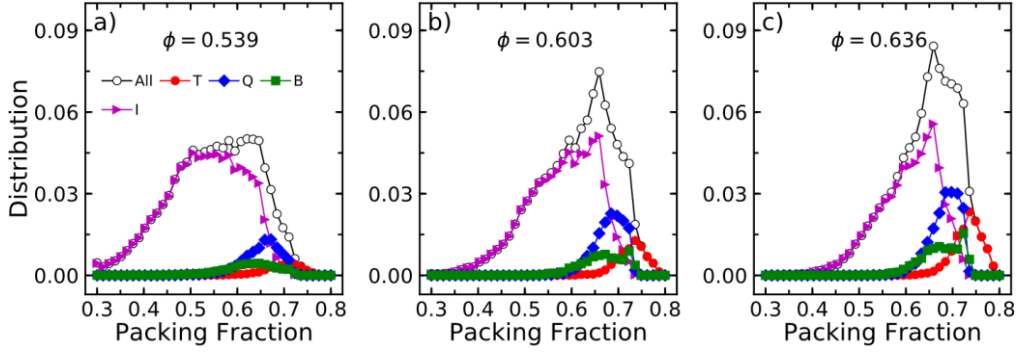

**FIG. 6.** Packing fraction distribution of tetrahedron (T), quartoctahedron (Q), BCC Simplex (B), irregular simplexes (I) and all simplexes in disordered packings at $\phi = 0.539$(a), $\phi = 0.603$(b), and $\phi = 0.636$(c) lying in the RLP line, respectively.

Note that the mean packing fraction of irregular simplexes exceeds 0.6 near $\phi_{RCP}$, which is very close to the value of B at RLP point, as shown in FIG. 5(a). Considering the two-order parameter theory [46], this implies that the density order in the disordered packings at $\phi_{RCP}$ may be ready for the formation of crystal phase. Thus, the disordered packing may be unstable and crystallizes at any time as packing fraction pass through $\phi_{RCP}$. This situation is quite similar to the Lennard-Jones fluid at large supercooling where the liquid crystallizes by a spinodal mechanism [47]. Another interesting phenomenon is that the mean packing fraction of regular simplexes (T, Q, and B) at RLP point is just around 0.64 (see FIG. 5(a)), coincident with $\phi_{RCP}$, which could have some implications on the critical density.

### IV. Discussion & conclusion

It should be noted that the volume fraction of T at $\phi \approx 0.64$ shown in FIG. 1(a) is much less than 30% reported in Refs. [21,24], where the difference of the maximal edge lengths $e_{max}$ from unit in Delaunay simplexes $\delta = e_{max} - 1$ and the condition of $\delta <$ 0.255 were used to select T. This is mainly because that the condition of $\delta < 0.255$ used to identify T from Delaunay simplexes is relatively loose, and some "quasi-regular tetrahedron" are also included [44,48]. With the condition of $\delta < 0.255$, it is difficult to make a reasonable justification for identifying T from $\delta$ distribution of Delaunay simplexes in disordered packings, as shown in FIG. 7. The boundary of $\delta = 0.255$ for T was initially found by Hale by trial in the proof of the Kepler conjecture [4], which might not be suitable for random packings.

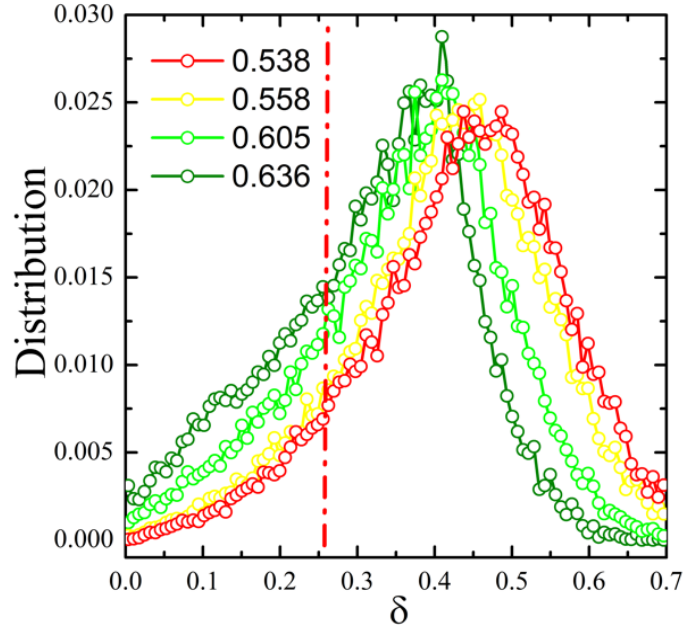

**FIG. 7.** $\delta$ distribution of the Delaunay simplexes in the disordered packings of $\phi = 0.538$, 0.558, 0.605, and 0.636, respectively. Here the shape of tetrahedra was evaluated according to $\delta = e_{max} - 1$, the difference of the maximal edge lengths $e_{max}$ from unit in Delaunay simplexes. The vertical red line indicates the boundary of $\delta = 0.255$ below which a simplex is regarded as regular tetrahedron in Ref. [2].

To accurately identify T from arbitrary Delaunay simplexes in random packings, one should also distinguish T from other regular simplexes, such as B. For example, T and B are quite similar in shape. If T and B cannot be effectively distinguished, the population of T will be overestimated. Unfortunately, the approach of $\delta = e_{max} - 1$ cannot be used to identify B. To estimate the overlapping of T identified by $\delta < 0.255$ with B, we calculated $\delta$ value for each simplex in the disordered packing at $\phi \approx 0.636$, and made comparisons with the square of Procrustean distance $d^2$ value of each simplex with respect to the template B. FIG. 8(a) shows the contour of $d^2(B) \sim \delta(T)$ for all simplexes in the packing at $\phi \approx 0.636$. The horizontal and vertical lines mark the boundary values of $d^2$ (=0.01) and $\delta$ (=0.255, 0.20, 0.15) for selecting B and T, respectively. It can be seen that if the boundary of $\delta = 0.255$ is applied, the overlapped T and B is about 75.5% of the total B. Even for the boundary of $\delta = 0.15$, the overlapped T and B is about 16.8% of the total B. Thus, the identified T by the condition of $\delta < 0.255$ include significant portion of B. Therefore, the population of both T and polytetrahedral aggregates in random packings was overestimated in previous studies [21,24]. If the boundary of $\delta = 0.15$ is applied, the fraction of the identified T is only about 13% at $\phi \approx 0.64$ [21,24]. We also analyzed the overlapping of T and B both identified by Procrustean distance approach. FIG. 8(b) shows the contour of $d^2(B) \sim d^2(T)$ for all simplexes in the packing at $\phi \approx 0.636$. The overlapping is only about 15.1% of the total B. Therefore, Procrustean distance approach is able to identify different regular simplexes and rigorous calibration has been employed in our work to maximally distinguish among T, Q and B. Thus, our analysis indicates that the configurations associated with crystal phases are much more important than the polytetrahedral ones in disordered packings.

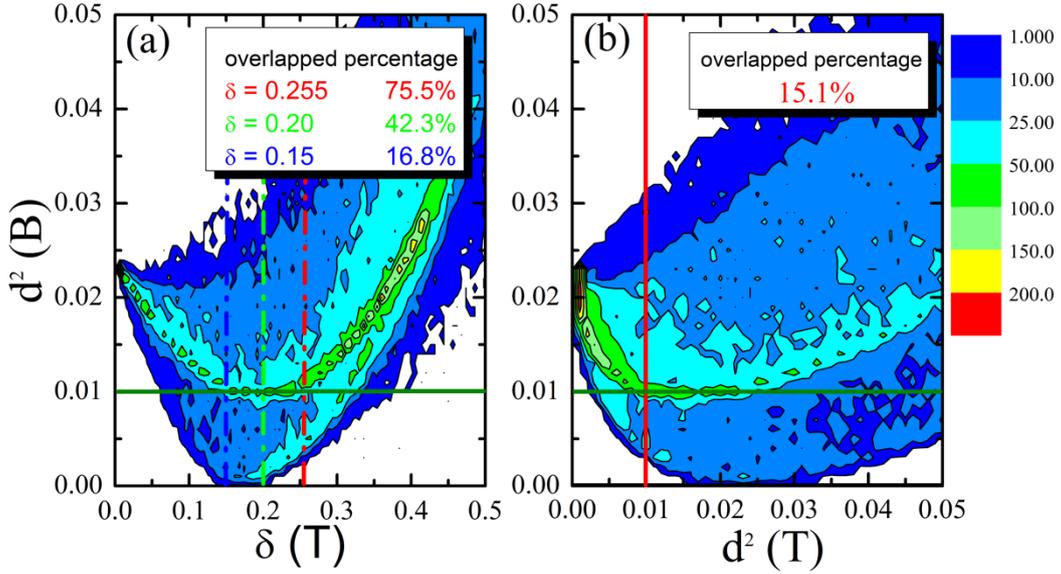

**FIG. 8. Comparison of the tetrahedra identified by Procrustean distance approach $d^2$ and the approach of $\delta = e_{max} - 1$.** The contours of $d^2(B) \sim \delta(T)$ (a) and $d^2(B) \sim d^2(T)$ (b) for all simplexes in the packing at $\phi \approx 0.636$. The horizontal and vertical lines in (a) mark the boundary values of $d^2$ and $\delta$ for selecting B and T, respectively. The horizontal and vertical lines in (b) mark the boundary values of $d^2$ for selecting B and T. About 16.8%, 42.3% and 75.5% of BCC simplexes cannot be distinguished from the tetrahedra identified by the condition of $\delta < 0.15$, 0.20 and 0.255, respectively. However, only about 15.1% of BCC simplexes is overlapped with the tetrahedra identified by the condition of $d^2 < 0.01$.

Note that lying in $\phi_{RCP}$, four packings with $Z \approx 4$ contain relatively higher population of the crystal-type and defective I clusters, compared to others with Z>4 (see FIG. 3(a) and (b)). This may result from the preparation of random packings. As mentioned in Ref.[33,43], the smaller the mechanical coordination number Z is, the more difficultly the mechanically stable random packings are prepared. This demonstrates that our structural model is very sensitive to the local ordering in random structures and able to efficiently detect it. In addition, our model can be applied to not only jammed random packings, but also other disordered sphere packings, even the packings beyond $\phi_{RCP}$, so that the disorder-to-order transition in hard sphere packings can be also explored by our model.

In conclusion, we developed a new tetrahedral structural model. It is revealed that random packings of monodisperse spheres possess intrinsic structure feature, which depends only on packing fraction. Instead of geometric frustrated structures, the local configurations associated with crystal phases are the essential features in disordered packings at the RCP line, which ensures that the crystal phases may naturally form as density passes through $\phi_{\text{RCP}}$. Our study provides the underlying structural basis for the random close packing.

**ACKNOWLEDGMENTS**

We are grateful to Dr. Y. L. Jin for valuable discussions. This work was supported by National Natural Science Foundation of China (Nos. 52031016 and 51631003).

**APPENDIX A: SIMULATION DETAILS OF RANDOM PACKINGS OF MONODISPERSE SPHERES**

In this study, the jammed random packings with different interparticle friction coefficients were investigated [33,43,49]. Different friction coefficients may produce different mechanical coordination numbers (Z) which characterizes the mechanical features at the grain scale in random packings. Each packing contains 10000 identical spheres in a cubic box with periodic boundary conditions. The details of the algorithm for preparation of random packings can be found in Refs. [33,43], according to which, the jammed random packings with mechanically stable and force-equilibrated at different packing fractions can be generated. All jammed random packings lie within a region in $Z - \phi$ plane demarcated by the random close packing (RCP) line, random loose packing (RLP) line and granular line, as shown in Ref.[33]. The RCP, RLP and granular lines correspond to the limit of vanishing compactivity, the limit of infinite compactivity, and the limit of infinite friction coefficient (Z=4), respectively.

**APPENDIX B: PROCRUSTEAN DISTANCE APPROACH: EVALUATION OF THE SHAPE OF A DELAUNAY SIMPLEX**

In this work, to determine all Delaunay simplexes of the packings, Voronoi tessellation method was employed to decompose the three-dimensional structures of random packings into Voronoi network by construction of bisecting planes along the lines joining the central spheres and all its neighbors. In the Voronoi network, each vertex is incident to four spheres which form a tetrahedron of a general shape, defining the Delaunay simplex. Thus, all Delaunay simplexes of the packings can be determined. They are the simplest elements in the three-dimensional structure.

There are several possible criteria to quantitatively evaluate the shape of an arbitrary Delaunay simplex. Here we employed the Procrustean distance (PD) approach, in which the proximity of an arbitrary Delaunay simplex to a given template can be quantitatively characterized by the square of Procrustean distance

$$d^2 = \min_{c,R,t,P} \left\{ \frac{1}{n} \sum_{i=1}^{n} \|x_i - (cRy_i + t)\|^2 \right\},$$

where $\{x_1, x_2, \cdots, x_n\}$ and $\{y_1, y_2, \cdots, y_n\}$ are the coordinates of $n$ spheres in a Delaunay simplex and the template, respectively. The minimum is calculated over all three-dimensional rotations $R$, translations $t$, size scaling $c$, and all possible combinatorial mapping between the two sets of coordinates *P*. From the definition, it is obvious that smaller values of *d* indicate more regular simplex to the template. Here, regular tetrahedron (T), quartoctahedron (Q) and BCC Simplex (B) serve as the templates. For an arbitrary Delaunay simplex, $d^2$ can be calculated for the template of T, Q, and B, respectively. FIG. 9 shows the $d^2$ distribution of Delaunay simplexes in random packings at different packing fraction to the template of T, Q, and B, respectively. PD distribution of the Delaunay simplexes are broad, and shifts to smaller distance with increasing packing fraction. In addition, the distribution at around zero distance increases, as shown in FIG. 9. These indicate that the simplexes in disordered packings become more and more regular, and more T, Q, and B are formed with the increase of packing fraction. We use $d^2$ as an order parameter to define the emergence of T, Q, or B in disordered packings as $d^2$ is smaller than a cutoff value.

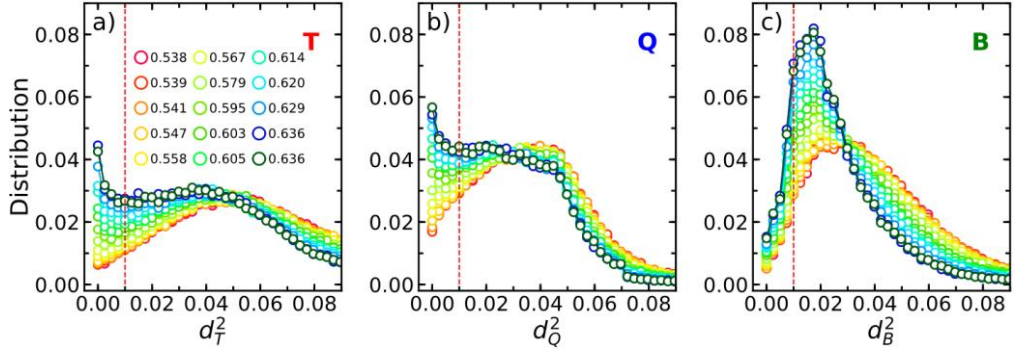

**FIG. 9.** Distribution of the square of Procrustean distance of Delaunay simplexes in disordered packings of monodisperse spheres on random loose packing (RLP) line (see FIG. S3) at different packing fractions to the template of tetrahedron (T) (a), quartoctahedron (Q) (b), and BCC simplex (B) (c), respectively. The vertical dashed lines mark the cutoff distances of $d_T^2=0.01$, $d_Q^2=0.01$, and $d_B^2=0.01$, respectively, which was determined in FIG. 10.

**APPENDIX C: DETERMINATION OF THE PROCRUSTEAN DISTANCE**

**CUTOFF**

In order to assign a Delaunay simplex to a given template shape, a cutoff distance needs to be specified for tetrahedron (T), quartoctahedron (Q), and BCC simplex (B), respectively, i.e., $d_T$, $d_Q$, and $d_B$, below which a simplex is considered to be a particular shape. To determine the Procrustean distance cutoff for identifying T, Q, and B, thermal effect on crystal structures was also taken into account in our calibration. First, we carried out molecular dynamics simulations to anneal FCC (Cu), HCP (Zr), and BCC (V) crystal structures at 300K for 1ns, respectively. Due to the thermal effect, the atoms in each crystal structure are vibrating around their equilibrium positions, so that the simplexes in each crystal structure deviate from their perfect shapes. The deviation should exhibit a distribution. Second, we performed the calculations of the Procrustean distance for the simplexes in the annealed crystal structures with respect to each template, and a distance distribution in an annealed crystal structure for each template can be obtained, which reflects the deviation of the simplexes.

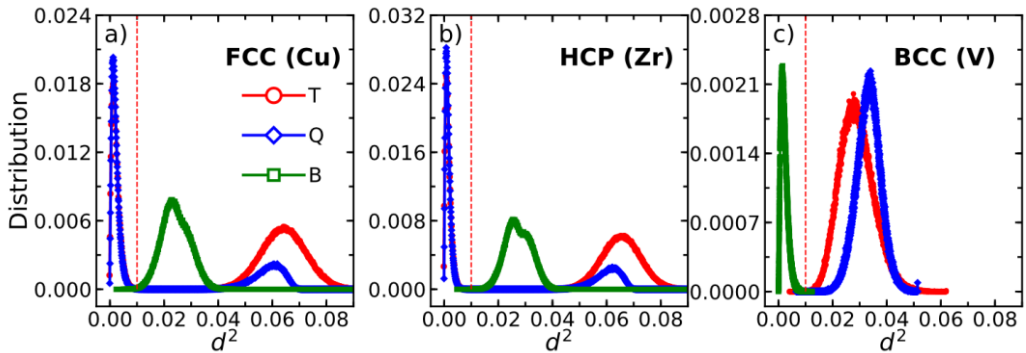

**FIG. 10.** The Procrustean distance distributions of Delaunay simplexes with respect to the template of T(red), Q(blue), and B(green) in FCC (a), HCP (b), and BCC (c) crystals annealed at 300K, respectively. The vertical dashed line in each panel indicates that the simplexes with $d^2 \leq 0.01$ can be regarded as T, Q, or B.

FIG. 10(a-c) shows the Procrustean distance distributions for the template of T, Q,

and B in the annealed FCC, HCP, and BCC crystal structures, respectively. To determine the best cutoff for T, Q, and B, the following criteria have to be taken into account simultaneously: (1) the cutoff should be as small as possible, so that more regular simplexes to the corresponding templates can be selected; (2) the cutoff should lead to the overlapping of T, Q, and B as small as possible, so that they can be maximally distinguished. Based on the above considerations, the cutoff of $d^2$=0.01 was chosen in the calibration of all Delaunay simplexes to the templates of T, Q, and B, respectively, as shown in FIG. 10. The vertical dashed lines indicate that the simplexes with $d^2 \leq 0.01$ can be regarded as T, Q, or B. For an arbitrary Delaunay simplex $i$, $d_i^2$ was calculated with respect to the template of T, Q, and B, respectively, so that $d_{iT}^2$, $d_{iQ}^2$, and $d_{iB}^2$ were obtained. If only one value is smaller than 0.01, the Delaunay simplex is assigned to the shape of the corresponding template. If there are two or three values are smaller than 0.01, the smallest one is chosen and the corresponding shape is assigned to the Delaunay simplex. If they are all larger than 0.01, the Delaunay simplex is classified into irregular simplex. Therefore, in our work, $d_T^2$=0.01, $d_Q^2$=0.01, and $d_B^2$=0.01 are employed in the calibration of all simplexes. All selected simplexes are regarded as regular simplexes. Simplexes with $d^2$>0.01 to all three templates are denoted as irregular simplexes.

# SUPPLEMENTARY INFORMATION FOR

## Intrinsic structural features in jammed disordered packing of monodisperse spheres

S. Q. Jiang and M. Z. Li*

*Department of Physics, Beijing Key Laboratory of Opto-electronic Functional Materials & Micro-nano Devices, Renmin University of China, Beijing 100872, China*

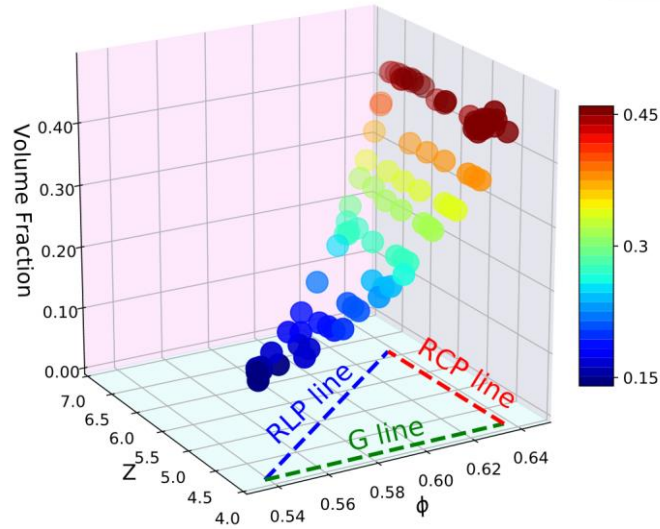

**FIG. S1.** The volume fraction of regular simplexes (T+Q+B) in disordered sphere packings at different packing fraction $\phi$ and mechanical coordination number Z. The color of a sphere represents the volume fraction indicated by the color bar. All jammed disordered packings lie within the triangle demarcated by the dashed lines in $Z - \phi$ plane which represent the random loose packing (RLP) line, random close packing (RCP) line, and granular line, respectively [1].

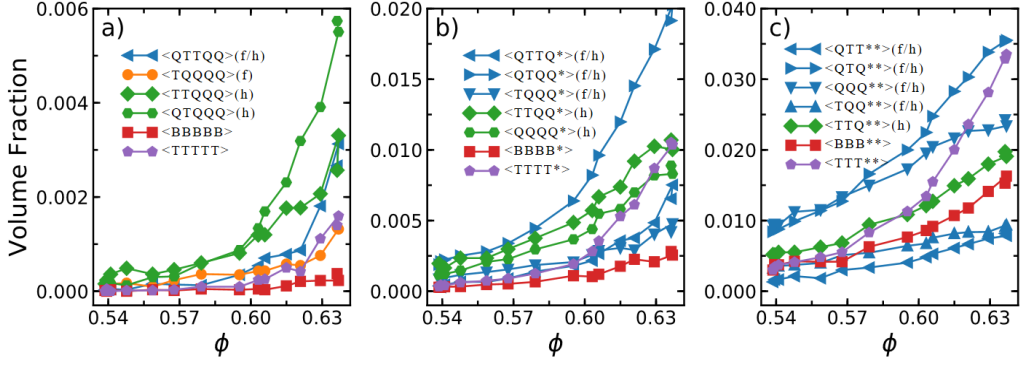

**FIG. S2.** Population of crystal-type clusters (a) and defective clusters (b, c) as a function of packing fraction in the packings lying in the RLP line, respectively. The population of the geometric frustrated clusters of <TTTTT> (a), <TTTT*> (b) and <TTT**> (c) was also presented with purple pentagons. The population of HCP-type clusters of <TTQQQ> and <QTQQQ> is even higher than the FCC-type clusters, indicating that HCP phase may be formed as disordered packings are crystallized. On the other hand, the population of <BBBBB> clusters is almost zero, although the fraction of B is similar to T as shown in FIG. 2. This implies that BCC crystal phase is hardly formed in disordered sphere packings, consistent with previous studies [2,3].